\documentclass[twocolumn,aps,pra]{revtex4}
\usepackage{epsfig}
\usepackage{amsmath}
\usepackage{amsfonts}
\usepackage{color}
\usepackage{soul}
\usepackage{multirow}
\usepackage{amssymb}

\def\e{\vec{\mathrm{e}}}
\def\FFF{\hat{\mathfrak{F}}}
\def\EE{\mathcal{E}}
\def\FF{\mathfrak{F}}
\def\III{\hat{\mathcal{I}}}
\def\QQ{\hat{Q}}

\begin{document}

\title{Magic radio-frequency dressing for trapped atomic microwave clocks}
\author{G.~A.~Kazakov\footnote{E--mail:
kazakov.george@gmail.com}, T.~Schumm}
\affiliation{{\setlength{\baselineskip}{18pt}
{Vienna Center for Quantum Science and Technology, Atominstitut, TU Wien, Stadionallee 2, 1020 Vienna, Austria}\\
}}

%%%%%%%%%%%%%%%%%%%%%%%%%%%%%%%%%%%%%%%%%%%%%%%%%%%%%%%%%%%%%%%%%%%%%%%%%%%%%%%%%%%%%%%%%
%%%%%%%%%%%%%%%%%%%%%%%%%%%%%%%%%%%%%%%%%%%%%%%%%%%%%%%%%%%%%%%%%%%%%%%%%%%%%%%%%%%%%%%%%
\begin{abstract}
It has been proposed to use magnetically trapped atomic ensembles to enhance the interrogation time in microwave clocks. To mitigate the perturbing effects of the magnetic trap, near-magic-field configurations are employed, where the involved clock transition becomes independent of the atoms potential energy to first order. Still, higher order effects are a dominating source for dephasing, limiting the perfomance of this approach. Here we propose a simple method to cancel the energy dependence to both, first and second order, using weak radio-frequency dressing. We give values for dressing frequencies, amplitudes, and trapping fields for $^{87}$Rb atoms and investigate quantitatively the robustness of these second-order-magic conditions to variations of the system parameters. We conclude that radio-frequency dressing can suppress field-induced dephasing by at least one order of magnitude for typical experimental parameters.
\end{abstract}

\pacs{???}

\maketitle                

%%%%%%%%%%%%%%%%%%%%%%%%%%%%%%%%%%%%%%%%%%%%%%%%%%%%%%%%%%%%%%%%%%%%%%%%%%%%%%%%%%%%%%%%%
%%%%%%%%%%%%%%%%%%%%%%%%%%%%%%%%%%%%%%%%%%%%%%%%%%%%%%%%%%%%%%%%%%%%%%%%%%%%%%%%%%%%%%%%%
\section{Introduction}
\label{sec:intro}

The performance of atomic clocks is closely linked to the interrogation time of the quantum oscillator. In microwave clocks, switching from thermal beams to atomic fountains has increased the interrogation time by about two orders of magnitudes, significantly improving the short-term stability. For example, the PTB CS2 primary beam standard with an interrogation time of about 8\,ms provides a short-term stability of $3.6\times 10^{-12}~\sqrt{\tau/1\,\mathrm{s}}$~\cite{Bauch03}. At the same time, the Cs fountain standard with an interrogation time of 0.8 \,s demonstrated a short-term stability of $4\times 10^{-14}~\sqrt{\tau/1\,\mathrm{s}}$~\cite{Santarelli99}, almost 2 orders of magnitude better.

In this spirit, it has been proposed to further enhance the interrogation time by working with trapped thermal atomic ensembles~\cite{Rosenbusch09}. Especially magnetically trapped alkali atoms on atom chips promise to combine long interrogation times with  fast and robust preparation and small system footprint and power consumption~{\cite{Lacroute10}.

In general, atomic microwave clocks rely on a measurement of the phase evolution of a superposition of two atomic ``clock'' states $|1\rangle$ and $|2\rangle$, usually implemented in the two hyperfine ground states of alkali atoms such as Cesium or Rubidium. Inhomogeneous external (trapping) fields lead to spatially varying energy shifts for the states $|1\rangle$ and $|2\rangle$ and hence to a position-dependent phase evolution. In a thermal atomic ensemble, this leads to dephasing, degrading the clock signal over time. This effect could be mitigated in ``magic traps'', where the energy shift for both states $|1\rangle$ and $|2\rangle$ is exactly identical, independent of the atoms position in the trap. 

So far, it is only possible to build ``near magic'' traps, where the non-equivalence of the trapping potental experienced by states $|1\rangle$ and $|2\rangle$ vanishes in the first order (in potential energy), but remains in higher orders, introducing a residual inhomogeneity into the system. An example is a static ({dc}) magnetic trap, where the atoms are confined in space, experiencing a local magnetic field $\vec{B}$ with a magnitude $B$  close to a so -called ``magic'' value $B_{magic}$. At this value, the relative energy shift $\Delta E$ between the states $|2\rangle$ and $|1\rangle$ features a minimum, however its second derivative remains non-zero. At finite temperature, atoms sample a distribution of fields $B$ different from $B_{magic}$, introducing dephasing.

In atomic systems where the atomic interactions are repulsive, like in $^{87}$Rb, the trap-induced energy shift can be partially compensated by the collisional shift, proportional to the atomic density~\cite{Rosenbusch09}. This method has been used in an atom chip clock based on trapped $^{87}$Rb atoms, where coherent interrogation over more than 2\,s could be demonstrated~\cite{Ramirez11}.

Here we propose to add the technique of magnetic radio-frequency (rf) dressing to selectively modify the potential landscape experienced by the two clock states in a static magnetic trap. rf dressing is a well-established method for the manipulation of ultra-cold atomic gases and Bose-Einstein condensates~\cite{Zobay01, Lesanovsky06}, commonly used for the generation of complex trapping geometries such a double-wells~\cite{Schumm05, Hofferberth07}, two-dimensional systems~\cite{Merloti13}, or ring topologies~\cite{Fernholz07}. In~\cite{Sinuco13} it has been pointed out that rf dressing can be used to modify the curvature of magnetic traps for $^{87}$Rb in a (hyperfine) state-dependent way. {In \cite{Zanon12} it was proposed to use rf dressing for the cancellation of first-order magnetic variations of the clock shift in optical clocks based on fermionic alkali-earth-like atoms.  Microwave dressing was used to reduce Rydberg atom susceptibility to varying dc electric field in \cite{Jones13}.}
%--------------------------------------------------------------------------------------
\begin{figure*}[ht]
\begin{center}
\resizebox{0.97\textwidth}{!}
{\includegraphics{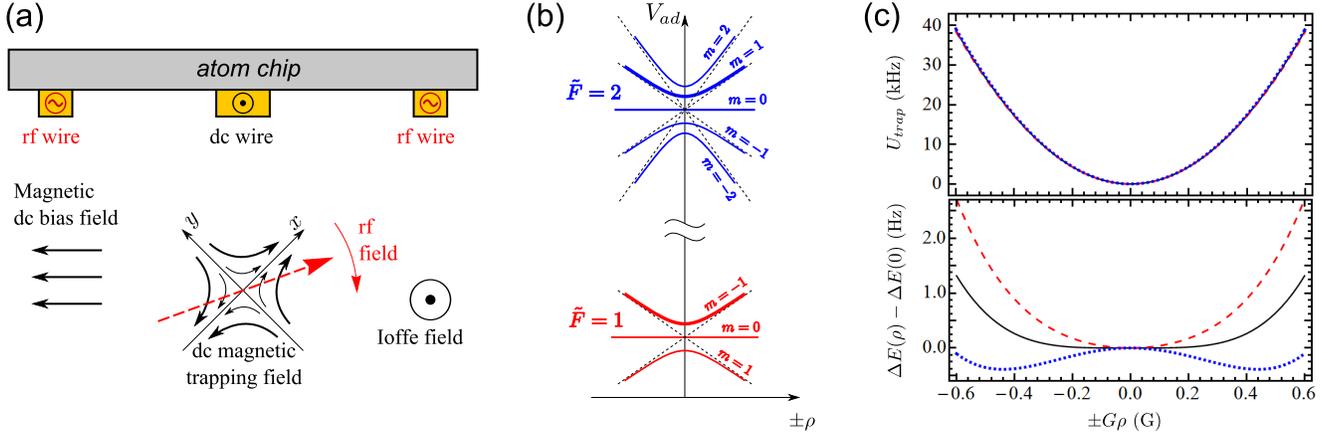}}
\end{center}
\caption{(Color online) (a) Schematic of a chip-based Ioffe-Pritchard trap with rf dressing (b) Spatial dependence of the adiabatic potential for different Zeeman states of the $F=1$ and $F=2$ ground state manifolds of $^{87}$Rb in the absence of rf dressing. The clock states $|1\rangle$ and $|2\rangle$ are indicated in bold. (c) Trapping potential $U_{trap}=V_{ad}(\rho)-V_{ad}(0)$ and relative energy shift $\Delta E$ as a function of position for zero rf field and 3 different values of Ioffe field: $B_I=B_{magic}$ (solid, black), $B_I=B_{magic}+0.03\,\mathrm{G}$ (dashed, red), $B_I=B_{magic}-0.03\,\mathrm{G}$ (dotted, blue).}
\label{fig:f1}
\end{figure*}
%---------------------------------------------------------------------------------------

In the present paper, we demonstrate that weak rf dressing can be used to elimination both, the first and second derivative of the relative energy shift between the states $|1\rangle$ and $|2\rangle$ with respect to the magnitude $B_0$ of the {dc} magnetic field in the trap. We refer to this as {\em second-order-magic conditions} in contrast to {\em first-order-magic conditions}, attainable in static magnetic traps, where only the first derivative of the relative energy shift vanishes. We identify and characterize these conditions for $^{87}$Rb atoms trapped in a rf dressed Ioffe-Pritchard-type trap, compare conventional {dc} first-order-magic Ioffe-Pritchard traps with second-order-magic traps, and characterize the robustness of this second-order magic potential to deviations of magnitude and polarization of the involved fields. {Note that also microwave dressing directly coupling atomic hyperfine levels can be used for suppression of both first- and second-order differential Zeeman shift in $^{87}$Rb, as demonstrated recently in \cite{Sarkany14}.}

%%%%%%%%%%%%%%%%%%%%%%%%%%%%%%%%%%%%%%%%%%%%%%%%%%%%%%%%%%%%%%%%%%%%%%%%%%%%%%%%%%%%%%%
\section{Physical model}
\label{sec:mod}
%+++++++++++++++++++++++++++++++++++++++++++++++++++++++++++++++++++++++++++++++++++++++
\subsection{Geometry and Hamiltonian}
\label{sec:mod1}

We consider the generic case of a magnetic Ioffe-Pritchard trap for $^{87}$Rb atoms. rf dressing can be conveniently implemented in atom chip setups using strong magnetic near fields, see Figure~\ref{fig:f1}(a). However, rf field amplitudes required for second-order magic conditions are weak (order 10-100\,mG) and can equally well be created by external coils~\cite{Merloti13}. For the sake of simplicity, we neglect gravity effects and a possible spatial inhomogeneity of the rf field. The static ({dc}) magnetic field $\vec{B}_0$ can be expressed as
%------------------------
\begin{equation}
\vec{B}_0=\e_z B_{I}+G(\e_x x-\e_y y). \label{eq:1}
\end{equation}
%------------------------
Near the trap axis $z$, the absolute value $B_0$ of this {dc} field is proportional to the square $x^2+y^2=\rho^2$ of the displacement from the axis, $B_0(\rho)\approx B_I+G^2\rho^2/(2 B_I)$.

The dressing radio-frequency field $\vec{B}_{rf}$ is equal to
%------------------------
\begin{equation}
\vec{B}_{rf}=\frac{B_{rf}}{2} \left[\left(\e_x \cos \delta - i \e_y \sin \delta\right)e^{i\omega t}+c.c. \right]\label{eq:2}
\end{equation}
%------------------------
where $\delta$ is a parameter characterizing the polarization of the rf field ($\delta=0,\, \pm \pi/4$ corresponds to linear and $\sigma^\pm$ circular polarization respectively). Although the parametrization (\ref{eq:2}) does not describe rf fields whose polarization ellipse axes are turned in the $(x,y)$ plane, it can describe any configuration of the local field up to rotations.

In the limit of a slowly moving atom, where the Larmor precession $\omega_L=\mu_BB/\hbar$ of the magnetic moment is much faster than the change of magnetic field in the rest frame of the atom, an adiabatic approximation becomes applicable: the atomic polarization follows the magnetic field adiabatically and the atom moves in a potential determined by the local characteristics of the magnetic fields only (see for example~\cite{Folman02} and references therein). The Hamiltonian governing the atomic dynamics is
%------------------------
\begin{equation}
\hat{H}^i=\frac{\hbar \omega_{hfs}}{2} \hat{\vec{J}} \cdot \hat{\vec{I}}+
\mu_B \left(g_J \hat{\vec{J}}+g_I \hat{\vec{I}}\, \right)\cdot
\left(\vec{B}_0+\vec{B}_{rf}(t)\right).  
\label{eq:3}
\end{equation}
%------------------------
Here $\hat{\vec{J}}$ and $\hat{\vec{I}}$ are the electronic shell and nuclear magnetic moments respectively (for the ground state of $^{87}$Rb, $J=1/2$, $I=3/2$),  $g_J=2.00233113$ and $g_I=-0.0009951414$~\cite{Steck08} are the corresponding gyromagnetic ratios, $\mu_B$ is the Bohr magneton, $\omega_{hfs}$ is the hyperfine splitting frequency, and the index ``$i$'' refers to ``initial''. In the absence of the rf dressing field, Hamiltonian (\ref{eq:3}) can be diagonalized analytically, yielding the well-known Breit-Rabi formula for the hyperfine energy spectrum~\cite{Steck08}:
%------------------------
\begin{eqnarray}
E^{BR}_{|\tilde{F}=I\pm J,m\rangle}&=&g_I\mu_B m B_0-\frac{\hbar \omega_{hfs}}{2 (2I+1)} \label{eq:4}\\
& \pm & \frac{\hbar \omega_{hfs}}{2}\sqrt{1+\frac{4 m X}{2 I+1}+X^2}\, , \nonumber
\end{eqnarray}
%------------------------
where
%------------------------
\begin{equation}
X=\frac{(g_J-g_I)\mu_B B_0}{\hbar \omega_{hfs}}. \label{eq:5}
\end{equation}
%------------------------
Eigenstates $|\tilde{F},m\rangle$ may be characterized by the projection $m$ of the total angular momentum $\hat{\vec{F}}$ on the magnetic field, and by the asymptotic value $\tilde{F}$ of the total angular momentum $F$. In the limit $B_0\rightarrow 0$, $F$ becomes a conserved quantity, and the eigenstates $|\tilde{F},m\rangle$ become states $|F,m\rangle$ with determined values $F$ of the total angular momentum. For $B_0\neq 0$, all the eigenstates $|\tilde{F},m\rangle$ except $|\tilde{F}=2,m=\pm 2\rangle$ contain both $|F=1,m\rangle$ and $|F=2,m\rangle$ states, but, if the magnetic field is weak ($\mu_B g_J B_0 \ll \hbar \omega_{hfs}$), the contribution of the state $|F\neq\tilde{F},m\rangle$ into the eigenstate $|\tilde{F},m\rangle$ occurs to be small. 

We define the relative energy shift $\Delta E$ as the difference between the adiabatic potentials $V_{ad}$ for the clock states $|1\rangle=|\tilde{F}=1,\tilde{m}_F=-1\rangle$ and $|2\rangle=|\tilde{F}=2,\tilde{m}_F=1\rangle$ with subtracted zero-field hyperfine splitting: $\Delta E = V_{ad,|2\rangle}-V_{ad,|1\rangle}-\hbar \omega_{hfs}$. In the purely static magnetic trap, this shift experiences a minimum at $B_0=B_{magic}=3.228917$\,G. The second derivative of $\Delta E$ around this minimum is about $\partial \Delta E/\partial B_0\approx 863\,\mathrm{Hz/G^2}$. At first order magic condition $B_I=B_{magic}$, $\Delta E(\rho)-\Delta E(0)$ is proportional to the fourth power of the distance $\rho$ from the trap axis, or to the second power of the atoms local potential energy (see Figure~\ref{fig:f1}(c)). 

Often it is reasonable to choose $B_I$ slightly below $B_{magic}$. It allows to obtain a more uniform distribution of $\Delta E$ over the thermal atomic cloud, see the Figure~\ref{fig:f1}(c). For the sake of clarity, we will compare different potentials with zero derivatives of the relative energy shift on the trap axis in this work.

Our aim is to state-selectively modify the trapping potential using an additional weak rf dressing field. Such dressing allows to design a trap, where not only the first but also the second derivative of $\Delta E$ with respect to the adiabatic potential (directly proportional to $B_{0}$ in purely static or weakly dressed traps) becomes zero (vanishing forth order dependence in distance $\rho$ from the trap axis). In such dressed potentials, the trap-induced dephasing can be significantly reduced compared to static {dc} field Ioffe-Pritchard traps.

In the presence of an oscillating rf field, it is possible either to apply the Floquet formalism~\cite{Shirley65} to the Hamiltonian (\ref{eq:3}) with static and rf magnetic fields given by (\ref{eq:1}) and (\ref{eq:2}) directly, or to transform the Hamiltonian to the rotating frame using a weak-field limit for $\vec{B}_{rf}$. Under the assumption that the rf field can be treated as classical, the Floquet formalism is equivalent to the fully quantized dressed-atom approach~\cite{Shirley65, Chu04} and it allows to perform high-precision calculations of the rf dressed levels for a wide range of parameters. The transformation to the rotating frame in the weak rf field limit allows either to use the rotating wave approximation (RWA), or to apply the Floquet formalism to the transformed Hamiltonian.

%+++++++++++++++++++++++++++++++++++++++++++++++++++++++++++++++++++++++++++++++++++++++
\subsection{Weak rf field limit and transformation to the rotating frame}
\label{sec:mod2}

We start from the Hamiltonian (\ref{eq:3}) and express it as
%----------------------------------------
\begin{equation}
\hat{H}^i=\hat{H}_{BR}+\mu_B \left(g_J \hat{\vec{J}}+g_I \hat{\vec{I}}\, \right)\cdot
\vec{B}_{rf}(t), \label{eq:6}
\end{equation} 
%----------------------------------------
where $\hat{H}_{BR}= \left(\hat{\vec{J}} \cdot \hat{\vec{I}}\, \right) \hbar\, \omega_{hfs}/2 + \mu_B \left(g_J \hat{\vec{J}}+g_I \hat{\vec{I}}\, \right)\cdot \vec{B}_0$ is time-independent and can be diagonalized. Eigenenergies of $\hat{H}_{BR}$ are given by the Breit-Rabi formula (\ref{eq:4}). We suppose that $\omega \ll \omega_{hfs}$ and $B_{rf}\ll B_0\ll \hbar \omega_{hfs}/\mu_B$. This allows us to neglect far off-resonant couplings of different hyperfine manifolds by the rf field, and to replace the exact matrix elements $\langle \tilde{F}, m| \hat{\vec{J}}\cdot \vec{B}_{rf} |\tilde{F}, m'\rangle$, $\langle \tilde{F}, m| \hat{\vec{I}}\cdot \vec{B}_{rf} |\tilde{F}, m'\rangle$ by their approximate values $\langle {F}, m| \hat{\vec{J}}\cdot \vec{B}_{rf} |{F}, m'\rangle$, $\langle F,m| \hat{\vec{I}}\cdot \vec{B}_{rf} |F,m'\rangle$. We can then represent the Hamiltonian~(\ref{eq:6}) as a sum of two Hamiltonians $\hat{H}^i_1$ and $\hat{H}^i_2$ operating in the subspaces $V_1$ and $V_2$ spanned by the sets of states $|\tilde{F}=1,m\rangle$ and $|\tilde{F}=2,m\rangle$ respectively:

%----------------------------------------
\begin{align}
&\hat{H}^i=\hat{H}^i_1+\hat{H}^i_2, \label{eq:7}\\
&\hat{H}^i_{\tilde{F}}=\sum_{m}|\tilde{F}m\rangle E^{BR}_{|\tilde{F},m\rangle}\langle\tilde{F},m| +\mu_B \vec{B}_{rf}  \nonumber \\
&\hphantom{aa}\cdot\sum_{m,m'}|\tilde{F},m\rangle g_{F=\tilde{F}} 
\langle F, m|\hat{\vec{F}} |F, m' \rangle
\langle\tilde{F}, m'|, \label{eq:8}
\end{align} 
%----------------------------------------
where $\tilde{F}=F$, and
%----------------------------------------
\begin{align}
g_F&=g_J\frac{F(F+1)-I(I+1)+J(J+1)}{2 F(F+1)}\nonumber \\
&+g_I\frac{F(F+1)+I(I+1)-J(J+1)}{2 F(F+1)}. \label{eq:9}
\end{align}
%----------------------------------------

Now we express the {dc} magnetic trapping field (\ref{eq:1}) as
%---------------------------------------
\begin{equation}
\vec{B}_0=\e_z B_I + \sqrt{\chi} \left(\e_x \cos(\alpha)+\e_y \sin(\alpha) \right), \label{eq:10}
\end{equation}
%---------------------------------------
where $x=\rho \cos \alpha$, $y=-\rho \sin \alpha$, and $\chi=G^2\rho^2$ is the square of the transverse ($x,y$-plane) component of the {dc} field. Near the trap axis, the trapping potential is proportional to $\chi$, see Figure~\ref{fig:f1}(b). To describe the local field, we change the coordinate system: let the new axis $z'$ be parallel to $\vec{B}_0$, the new axis $x'$ lies in the plane $(\e_z,\vec{B}_0)$, and the new axis $y'$ shall be orthogonal to $x', z'$. Then, after some algebra, we express the rf field (\ref{eq:2}) as
%---------------------------------------
\begin{equation}
\vec{B}_{rf}=\frac{e^{i\omega t}}{2} \left[\e_{x'} B_{x'} - i \e_{y'} B_{y'} + \e_{z'} B_{z'}  \right] + c.c., \label{eq:11}
\end{equation}
%---------------------------------------
where $\e_{x'}$, $\e_{y'}$ and $\e_{z'}$ are the basis vectors of the new axes,
%---------------------------------------
\begin{align}
B_{x'}&=B_{rf} \left( \cos \alpha \, \cos \theta \, \cos \delta - i \sin \alpha \cos \theta \sin \delta \right), \nonumber \\
B_{y'}&=B_{rf} \left( \cos \alpha \sin \delta - i \sin \alpha \cos \delta \right), \label{eq:12} \\
B_{z'}&=B_{rf} \left( \cos \alpha \, \sin \theta \, \cos \delta - i \sin \alpha \sin \theta \sin \delta \right), \nonumber
\end{align}
%---------------------------------------
and $\theta$ is an angle between the trap axis $z$ and the direction of the {dc} field $\vec{B}_0$.

As a next step we apply a unitary transformation $\hat{U}_R=\exp\left[i(\hat{P}_{\tilde{1}}-\hat{P}_{\tilde{2}}) \hat{F}_{z'} \omega t \right]$ to transform the Hamiltonian into the frame rotating with angular velocity $\omega$ around the local direction of the static magnetic field~\cite{Lesanovsky06}. Here $\hat{P}_{\tilde{F}}$ is a projector onto the subspaces $V_{\tilde{F}}$. This yields the new Hamiltonian
%----------------------------------------
\begin{equation}
\hat{H}=\hat{U}_R^+\hat{H}^i\hat{U}_R-i\hbar\hat{U}_R^+\left(\frac{\partial \hat{U}_R}{\partial t}\right)=\hat{H}_1+\hat{H}_2, \label{eq:13}\\
\end{equation}
%----------------------------------------
where Hamiltonians $\hat{H}_{\tilde{F}}$ ($\tilde{F}=1,2$), in turn, may be represented as
%----------------------------------------
\begin{equation}
\hat{H}_{\tilde{F}}=\sum_{n=-2}^2\hat{H}^{(n)}_{\tilde{F}}\exp(in\omega t). \label{eq:14}
\end{equation}
%----------------------------------------
The Fourier components of these Hamiltonians are equal to
%----------------------------------------
\begin{widetext}
\begin{align}
\hat{H}_{\tilde{F}}^{(0)}&=\sum_{m}|\tilde{F},m\rangle 
\left(E^{BR}_{|\tilde{F},m\rangle}\pm \hbar \omega m_F  \ \right) \langle\tilde{F},m| +\frac{\mu_B g_F}{4} \left( \hat{F}_{\pm} (B_{x'}\mp B_{y'})+\hat{F}_{\mp} (B_{x'}^*\mp B_{y'}^*) \right), \label{eq:15} \\
\hat{H}_{\tilde{F}}^{(1)}&=\frac{\mu_B g_F}{2} B_{z'}\hat{F}_{z'}, \quad 
\hat{H}_{\tilde{F}}^{(2)}=\frac{\mu_B g_F}{4} \hat{F_\mp}(B_{x'}\mp B_{y'}), \quad 
\hat{H}_{\tilde{F}}^{(-1)}=\hat{H}_{\tilde{F}}^{(1)+}, \quad 
\hat{H}_{\tilde{F}}^{(-2)}=\hat{H}_{\tilde{F}}^{(2)+}.
\label{eq:16}
\end{align}
\end{widetext}
%----------------------------------------
Here the upper signs correspond to $\tilde{F}=1$, the lower ones correspond to $\tilde{F}=2$, and $\hat{F}_{\pm}=\hat{F}_{x'}\pm i \hat{F}_{y'}$.

Within the rotating wave approximation, one retains only $\hat{H}_{\tilde{F}}^{(0)}$. Also, it is possible to construct a Floquet Hamiltonian using rapidly oscillating terms. Such a combined weak-field Floquet approximation (WFFA) is more precise than the pure RWA. Also, the WFFA allows to classify the quasienergy spectrum in a more convenient way than it is possible in a straightforward Floquet analysis based on the Hamiltonian~(\ref{eq:6}), see Appendix for details. The WFFA representation furthermore simplifies the numerical algorithms to search for the second-order magic conditions.

%%%%%%%%%%%%%%%%%%%%%%%%%%%%%%%%%%%%%%%%%%%%%%%%%%%%%%%%%%%%%%%%%%%%%%%%%%%%%%%%%%%%%%%%
\section{Second-order magic conditions}
\label{sec:som}
If the rf field is absent or weak and far from resonances (referring to $\hbar\omega=|g_F|\mu_{B}B_{I}\approx\mu_{B}B_{I}/2$), the trapping potential in the Ioffe-Pritchard trap is proportional to the {dc} field magnitude $B_0$. Near the trap axis $z$, $B_0=\sqrt{B_I^2+\chi}\simeq B_I+\chi/(2 B_I)$, i.e. the trapping potential is proportional to $\chi$, see Figure~\ref{fig:f1}(c). The relative energy shift $\Delta E$ depends on $\chi$ as
%---------------------------------
\begin{equation}
\Delta E(\chi)=A_0+A_1\chi+A_2\chi^2+A_{3}\chi^3+...
\label{e:17}
\end{equation}
%---------------------------------
(the coefficients $A_i$ can have an angular dependence, if the rf field polarization is not perfectly circular, see Section~\ref{sec:rob2} for details). In a purely static first-order magic trap, $A_{1}$ vanishes for $B_I=B_{magic}$. Other coefficients are
%---------------------------------
\begin{align}
&A_{0}\approx -4497.4\,\mathrm{Hz}, \nonumber \\
%&A_{1}=0, \nonumber \\
&A_2=\frac{1}{2}\, \frac{1}{4 B_{magic}^2}\left.\frac{\partial^2 \Delta E}{\partial B^2}\right|_{B=B_{magic}} \approx \, \, 10.34\,\mathrm{Hz/G^4}, \nonumber \\
&A_3\approx  -0.49\,\mathrm{Hz/G^6}. \nonumber
\end{align}
%---------------------------------
Under second-order magic conditions, both $A_{1}$ and $A_{2}$ vanish, and the potential close to the trap axis can be characterized by the coefficients $A_0$ (indicating the absolute shift at the trap center) and $A_3$ (relevant for a remaining position-dependent dephasing). 

%+++++++++++++++++++++++++++++++++++++++++++++++++++++++++++++++++++++++++++++++++++++++
\subsection{Qualitative considerations}
\label{sec:som1}

To understand how rf dressing can mitigate the position-dependent dephasing in a Ioffe-Pritchard trap, we consider the following simplified model. We limit our considerations to the rotating wave approximation, where the system is described by the Hamiltonian~(\ref{eq:15}), and suppose that the atom is kept in the vicinity of the trap axis, where $G\rho \ll B_I$. The angle $\theta$ between the trap axis and the {dc} field direction is close to 0, so we set $\theta=0,~\alpha=0$ in (\ref{eq:12}).
%--------------------------------------------------------------------------------------
\begin{figure}
\begin{center}
\resizebox{0.4\textwidth}{!}
{\includegraphics{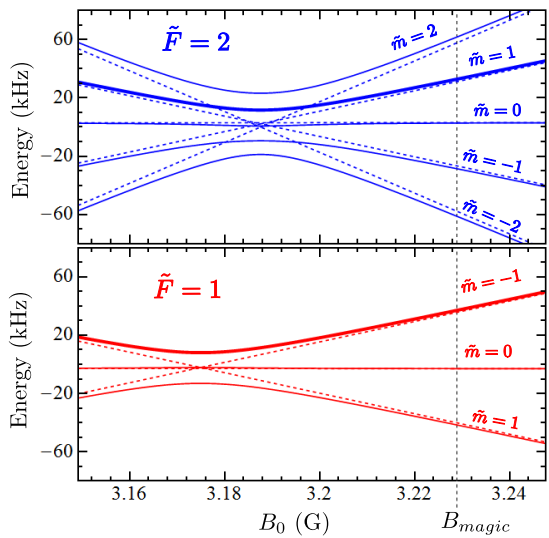}}
\end{center}
\caption{(Color online) Zeeman shifts of the upper (top) and lower (bottom) dressed manifolds of the ground state of $^{87}$Rb as given by Hamiltonian~(\ref{eq:15}) (zero-field hyperfine interaction terms are subtracted). Solid: dressed potentials, $\omega/(2 \pi)=2.23$~MHz, $B_{rf}=0.05$~G, $\delta=0$. Dashed: bare states. The clock states $|1\rangle$ and $|2\rangle$ are indicated in bold.}
\label{fig:f2}
\end{figure}
%---------------------------------------------------------------------------------------
Consider the energy shift $\Delta E$ as a function of $B_0$. Without rf dressing, it exhibits a minimum of $\Delta E\approx-4497.37$~Hz at $B_0=B_{magic}$ (see Figure~\ref{fig:f1}(c)). This function is convex, the second derivative $\partial^2 \Delta E/\partial B_0^2\approx 863\,\mathrm{Hz/G^2}$.  In consequence, a static Ioffe-Pritchard trap with $B_I<B_{magic}$ provides a slightly higher confinement (higher trap frequency) to the state $|1\rangle$ compared to state $|2\rangle$.

Adding an rf dressing field with a frequency below resonance ({\em low-frequency case}, $\hbar\omega < |g_F| \mu_{B}B_{I} \approx \mu_{B}B_{I}/2$}, as shown in Figure~\ref{fig:f2} adds a second convex contribution (with a minimum at resonance) to the energy shift of the weak-field seeking states $|1\rangle$ and $|2\rangle$. The curvature of this contribution depends on the state and the polarization of the rf dressing field. In Figure~\ref{fig:f2}, the rf dressing field is linearly polarized ($\delta=0$). One can see that the contributions to both levels $|1\rangle$ and $|2\rangle$ are essentially the same.
 
%--------------------------------------------------------------------------------------
\begin{figure}[b]
\begin{center}
%\resizebox{0.24\textwidth}{!}{\includegraphics{fig3a}}\resizebox{0.23\textwidth}{!}{\includegraphics{fig3b}}\\
\resizebox{0.45\textwidth}{!}{\includegraphics{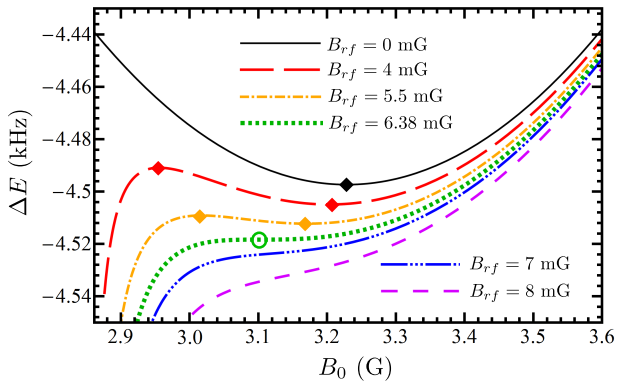}}
\end{center}
\caption{Dependences of the relative energy shift on the magnitude $B_0$ of the {dc} field for different magnitudes $B_{rf}$ of left-hand circularly polarized ($\delta=-\pi/4$) rf field. Here $\omega=2\pi\cdot 2$~MHz, diamonds denote the points where $\partial\Delta E/\partial B_0=0$, open circle denote the points where both $\partial\Delta E/\partial B_0=0$ and $\partial^2\Delta E/\partial B^2_0=0$. Calculations were performed in RWA.}
\label{fig:3}
\end{figure}
%---------------------------------------------------------------------------------------

By applying elliptically polarized rf fields, we can add more or less of this convex contribution to the energy of $|1\rangle$ compared to the energy of $|2\rangle$, implementing state-dependent dressing. If the rf field is left-handed polarized ($\delta=-\pi/4$), as shown in Figure~\ref{fig:f1}(a), only the $\tilde{F}=1$ manifold will be dressed, as follows from (\ref{eq:12}) and (\ref{eq:15}).  Figure~\ref{fig:3} illustrates the modification of $\Delta E$ in this case: the minimum of $\Delta E$ moves left (to lower $B_0$), and the second derivative $\partial^2 \Delta E/\partial B_0^2$ at the position of the minimum decreases. At second-order magic conditions, $\Delta E$ shows a saddle point (dotted green line in Figure~\ref{fig:3}). Note also, that the local field configuration in the Ioffe-Pritchard trap dressed by the circularly polarized rf field remains invariant with respect to rotations around the trap axis; the trapping potential remains axially symmetric. 

Similar considerations may be performed for the {\em high-frequency case}, when the rf frequency is above resonance ($\hbar\omega > |g_F|\mu_{B}B_{I}$, to the ``right side'' of $B_{magic}$ in Figure~\ref{fig:f2}). Then the weak-field seeking states lie on the lowest and 2nd lowest branches of the $\tilde{F}=1$ and $\tilde{F}=2$ manifolds respectively, the dressing leads to a decreasing concave contribution to the energy shift. To decrease the second derivative $\partial^2 \Delta E/\partial B_0^2$, one must use a right-hand ($\delta=\pi/4$) circular polarized rf field. However, these states become high-field seekers for atoms that are far from the trap axis, where the {dc} field $B_0$ becomes higher than the resonance value, and the trap becomes unstable (this situation resembles evaporative cooling in static magnetic traps). In this work we hence restrict our study to the low-frequency case.

%+++++++++++++++++++++++++++++++++++++++++++++++++++++++++++++++++++++++++++++++++++++++
\subsection{Results of numerical optimization}
\label{sec:som2}

%--------------------------------------------------------------------------------------
\begin{figure}
\begin{center}
%\resizebox{0.24\textwidth}{!}{\includegraphics{fig3a}}\resizebox{0.23\textwidth}{!}{\includegraphics{fig3b}}\\
\resizebox{0.47\textwidth}{!}{\includegraphics{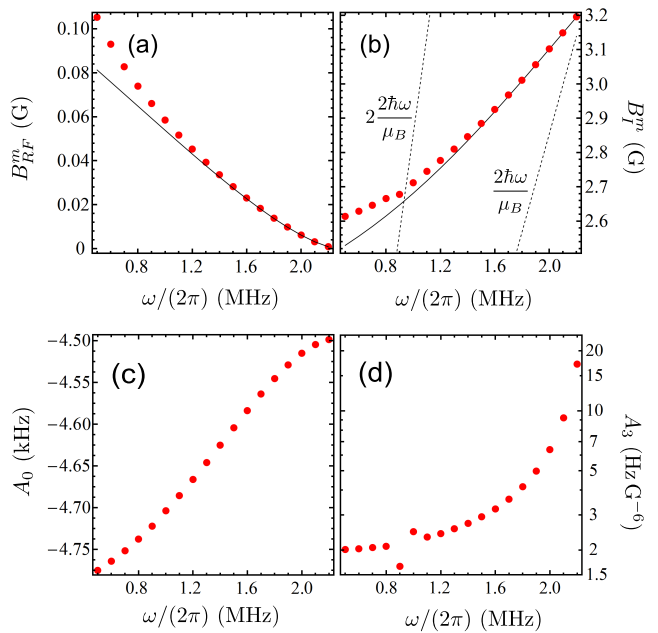}}
\end{center}
%\resizebox{0.43\textwidth}{!}{\includegraphics{fig3c}}\hphantom{a}
\caption{``Second-order magic'' magnitudes of rf dressing field (a) and Ioffe field (b) corresponding to different rf frequencies $\omega$, calculated in RWA (black solid curves), and WFFA with 21 manifolds (red circles). The dashed lines in plot (b) correspond to the two- and single-photon resonant conditions. (c) and (d) are the coefficients $A_0$ and $A_3$ in the expansion~(\ref{e:17}) for the ``second-order magic'' conditions calculated in WFFA.}
\label{fig:4}
\end{figure}
%---------------------------------------------------------------------------------------

Strictly speaking, the angle $\theta$ between the axis $z$ of Ioffe-Pritchard trap and the direction of the {dc} field $\vec{B}_0$ is equal to zero only on the axis $z$. Therefore, a simultaneous elimination of the derivatives $\partial \Delta E/\partial B_0$ and $\partial^2 \Delta E/\partial B_0^2$ of the energy shift $\Delta E$ with respect to the {dc} field magnitude $B_0$ at $\theta=0$ considered in Section~\ref{sec:mod1} is not exactly equivalent to the second-order magic conditions, i.e. simultaneous elimination of the derivatives $\partial \Delta E/\partial \chi$ and $\partial^2 \Delta E/\partial \chi^2$, although the qualitative analysis remains similar.

In this section we present values for Ioffe fields $B_I^m$ and rf field amplitudes $B_{rf}^m$ corresponding to the ``second-order magic'' conditions for different frequencies of the rf dressing field, calculated both in RWA based on the Hamiltonian~(\ref{eq:15}), and in WFFA based on the Hamiltonian~(\ref{eq:14}), see Figure~\ref{fig:4}(a,b) and Table~\ref{tab:1}.
%--------------------------------------------------------------------------------------
\begin{table}
\begin{center}
\caption{``Second-order magic'' magnitudes of rf dressing field and Ioffe field,  calculated according RWA and combined weak field-Floquet approach (WFFA).}
\begin{tabular}{|c|cc|cc|}\hline \hline
\multirow{2}{*}{  \hphantom{a}$\displaystyle{\frac{\omega}{2\pi}}$,~MHz\hphantom{a} }
&\multicolumn{2}{|c|}{RWA}&\multicolumn{2}{|c|}{Floquet}\\
& \hphantom{a}  $B^m_I$,~G \hphantom{a} & \hphantom{a} $B^m_{rf}$,~G \hphantom{a} & \hphantom{a}  $B^m_I$,~G \hphantom{a} & \hphantom{a} $B^m_{rf}$,~G \hphantom{a} \\ \hline
0.5 & 2.530 & 0.0813    & 2.614 & 0.1053  \\
0.6 & 2.556 & 0.0758    & 2.629 & 0.0931  \\
0.7 & 2.585 & 0.0704    & 2.646 & 0.0828  \\
0.8 & 2.615 & 0.0648    & 2.665 & 0.0739  \\
0.9 & 2.647 & 0.0593    & 2.678 & 0.0661  \\
1.0 & 2.681 & 0.0539    & 2.712 & 0.0585  \\
1.1 & 2.717 & 0.0484    & 2.745 & 0.0517  \\
1.2 & 2.755 & 0.0430    & 2.777 & 0.0453  \\
1.3 & 2.794 & 0.0377    & 2.810 & 0.0393  \\
1.4 & 2.834 & 0.0326    & 2.846 & 0.0336  \\
1.5 & 2.876 & 0.0275    & 2.885 & 0.0282  \\
1.6 & 2.920 & 0.0227    & 2.925 & 0.0231  \\
1.7 & 2.964 & 0.0181    & 2.967 & 0.0183  \\
1.8 & 3.009 & 0.0137    & 3.011 & 0.0138  \\
1.9 & 3.055 & 0.00971   & 3.056 & 0.00976 \\
2.0 & 3.102 & 0.00613   & 3.102 & 0.00615 \\
2.1 & 3.149 & 0.00310   & 3.149 & 0.00310 \\
2.2 & 3.195 & 0.000816  & 3.195 & 0.000816 \\
\hline \hline
\end{tabular}\label{tab:1}
\end{center}
\end{table}
%---------------------------------------------------------------------------------------
In the WFFA, the infinite Floquet ``Hamiltonian'' was truncated to $21\times 21$ matrix blocks. Pairs of $(B_I^m,B_{rf}^m)$ obtained in the WFFA were tested using a straightforward Floquet analysis based on the Hamiltonian~(\ref{eq:6}), and both $\partial \Delta E/\partial \chi$ and $\partial^2 \Delta E/\partial \chi^2$ remains zero up to the level of about 0.1\,$\mathrm{Hz/G^2}$ and below a few $\mathrm{Hz/G^4}$ respectively, which corresponds to a relative error in the determination of $B^m_I$ and $B^m_{rf}$ of about $0.1\,\%$. Note the appearance of a ``kink'' in the plot of $B_I^m$ near $\omega=2\pi\times 0.9$\,MHz caused by a distortion of the energy levels near a two-photon resonance, (when $|g_F| \mu_B B_I^m$ approaches $2\hbar\omega$). In Figure~\ref{fig:4}(c) and (d) we present the coefficients $A_0$ and $A_3$ of the expansion~(\ref{e:17}) for second-order magic conditions corresponding to different rf frequencies.

We see that the closer the rf frequency approaches the single-photon resonance condition, the weaker the rf field (amplitude) that should be applied to attain second-order magic conditions, and the rotating wave approximation becomes more precise. However, the coefficient $A_3$ also increases with rf frequency, and hence the residual position-dependent decoherence rate will also increases. Optimal parameters will depend on the density and temperature of the atomic ensemble.

%--------------------------------------------------------------------------------------
\begin{figure}
\begin{center}
%\resizebox{0.47\textwidth}{!}{\includegraphics{fig4a}}
\resizebox{0.45\textwidth}{!}{\includegraphics{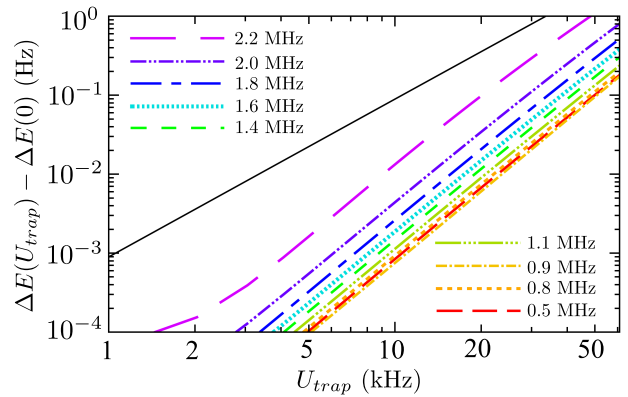}}
\end{center}
\caption{(color online) Relative energy shift for {dc} first-order magic trap (black solid curve), and for rf dressed second-order magic trap (coloured non-continuous curves) as a function of $U_{trap}=V_{ad}(\rho)-V_{ad}(0)$. Frequencies of the rf dressing fields are given in the legend, the corresponding values of $B_{rf}^m$ and $B_{I}^m$ can be found in Table~\ref{tab:1}.}
\label{fig:5}
\end{figure}
%---------------------------------------------------------------------------------------

To illustrate how the rf dressing improves the trap, we compare profiles of the relative energy shift $\Delta E-\Delta E(0)$ as a function of the trapping potential $U_{trap}(\chi)=V_{ad}(\chi)-V_{ad}(0)$ for an undressed ``first-order magic'' trap and for ``second-order magic'' traps corresponding to different rf field frequencies in Figure~\ref{fig:5}. 

One can see that for atomic ensembles cooled to temperature of the order of 1\,$\mu$K (about 20\,kHz in frequency units), the variation of $\Delta E$ over the dressed trap can be reduced by almost 2 orders of magnitude compared to the {dc} undressed magic trap. Further cooling will lead to an even stronger suppression of the position-dependent decoherence rate, because at such low energies, the relative energy shift is determined by the lowest order term of the Taylor expansion, proportional to $U_{trap}^3$ for dressed trap and to $U_{trap}^2$ for the undressed trap.

{Finally, we consider the question of validity of the adiabatic apporoximation near the resonance $\omega = \mu_B B_{magic}|g_F|/\hbar$. The adiabatic approximation is applicable, when the rate of change of the splitting $\EE$ of energy levels remains much less than the splitting itself: 
$
\frac{d \EE }{d\rho}\frac{d\rho}{dt} \ll \frac{\EE^2}{\hbar}.
$
Taking $\EE=\mu_B |g_1| B_0-\hbar \omega$ and estimating $d\rho/dt$ as $v_T=\sqrt{2 k_B T/m_{Rb}}$, we obtain that the adiabatic approximation is valid, when 
$
(\mu_B |g_F| B_I^m-\hbar \omega)^2 \gg 2\, k_B T \, \hbar \omega_{xy}.
$ 
Here we express $G$ via the transversal oscillation frequency $\omega_{xy}$ of the trap and estimate $\rho$ as $\sqrt{2 k_B T/(m_{Rb}\omega_{xy}^2)}$. Near the resonance, $(\mu_B |g_F| B_I^m-\hbar \omega)\approx 0.67(\mu_B |g_F| B_{magic}-\hbar \omega)$, see Figure \ref{fig:4} (b). The validity condition for the adiabatic approximation can hence be written as
$$
(\mu_B |g_F| B_{magic}/\hbar - \omega)^2 \gg 3\, k_B T \, \omega_{xy}/\hbar,
$$
where $\mu_B |g_F| B_{magic}/\hbar\approx 2.26$~MHz. As an example, we consider $^{87}$Rb atoms cooled down to 1~$\mu$K and confined in a trap with $\omega_{xy}=2\pi \times 2$~kHz. Then $\sqrt{3 k_{B}T \omega_{xy}/\hbar}\approx 2\pi\times 11$~kHz, and the adiabatic approximation is valid, if $2.26~\mathrm{MHz}-2\pi\omega \gg 11$~kHz. Less tight traps (relevant for atomic clocks because of a lower atomic number density and collisional shift) and colder atomic ensembles allow to approach even closer the resonance without loosing the validity of the adiabatic approximation.}
%--------------------------------------------------------------------------------------
\begin{figure}
\begin{center}
\resizebox{0.45\textwidth}{!}{\includegraphics{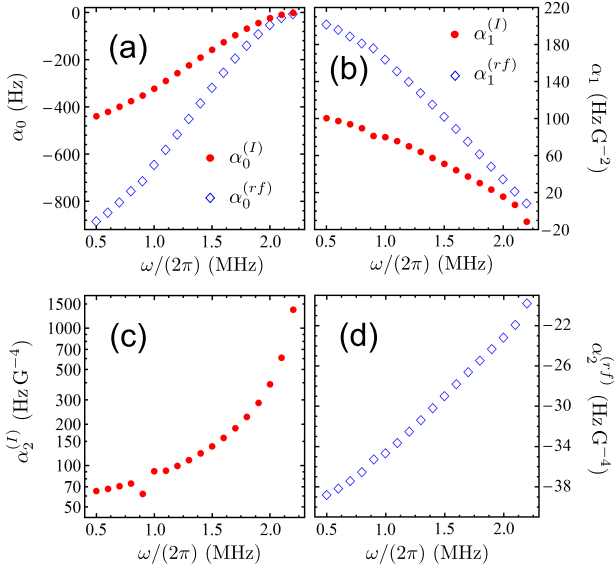}}
\end{center}
\caption{(color online) Parameters $\alpha_0^{I}$ and $\alpha_0^{rf}$ (a), $\alpha_1^{I}$ and $\alpha_1^{rf}$ (b), $\alpha_2^{I}$ (c) and $\alpha_2^{rf}$ (d), characterizing the robustness of the relative energy shift $\Delta E$ with respect to variations of the magnitude of the Ioffe field and the rf dressing field, see expansion (\ref{eq:18}).}
\label{fig:6}
\end{figure}
%---------------------------------------------------------------------------------------
%%%%%%%%%%%%%%%%%%%%%%%%%%%%%%%%%%%%%%%%%%%%%%%%%%%%%%%%%%%%%%%%%%%%%%%%%%%%%%%%%%%%%%%%
%%%%%%%%%%%%%%%%%%%%%%%%%%%%%%%%%%%%%%%%%%%%%%%%%%%%%%%%%%%%%%%%%%%%%%%%%%%%%%%%%%%%%%%%
\section{Robustness}
\label{sec:rob}
In any physical implementation of the dressed trap, magnitudes and polarizations of the involved fields can be controlled up to a certain accuracy only. These uncertainties must be taken into account for the proper development of the trap. Note that the pure {dc} first-order magic trap has a significant advantage, because at $B_I=B_{magic}$, the deviation of the relative energy shift $\Delta E$ is proportional to the squared deviation $\delta B_I$ of the Ioffe field $B_I$, namely $\Delta E(B_{magic}+\delta B_I)-\Delta E(B_{magic})=C \delta B_I^2$, where $C=431~\mathrm{Hz/G^2}$. The deviation of $\Delta E$ vanishes in the first order in $\delta B_I$.

In the following section we study the sensitivity of $\Delta E$ to deviations of $B_I$ and $B_{rf}$ from their second-order magic values, and to a deviation of the rf field polarization from the perfect left-hand circular one.

%+++++++++++++++++++++++++++++++++++++++++++++++++++++++++++++++++++++++++++++++++++++++
\subsection{Robustness to variations of field magnitudes}
Under second-order magic conditions, the coefficients $A_1$ and $A_2$ in the expansion~(\ref{e:17}) are equal to zero. If $B_I$ or $B_{rf}$ deviate from their ``magic'' values $B_I^m$ and $B_{rf}^m$ by $\delta B_I$ and $\delta B_{rf}$ respectively, this cause a change of all coefficients $A_i$. We expand these coefficients near the point $(B_{rf}^m,B_I^m)$  as
%----------------------------------------------
\begin{align}
A_i(B_I,B_{rf})&=A_i+\alpha^{(I)}_i  \frac{\delta B_I}{B_I} +\alpha^{(rf)}_i \frac{\delta B_{rf}}{B_{rf}}+... \label{eq:18}
\end{align}
%----------------------------------------------
where $\delta B_I=B_I-B_I^m$, $\delta B_{rf}=B_{rf}-B_{rf}^m$, and $A_i=A_i(B_I^{m},B_{rf}^m)$. Such a representation is convenient, as in many physical implementations the fields can be controlled to a known relative precision. The sensitivity to field fluctuations is expressed by the coefficients $\alpha_0$, $\alpha_1$ and $\alpha_2$; they are calculated in WFFA and represented in Figure~\ref{fig:6}.

%+++++++++++++++++++++++++++++++++++++++++++++++++++++++++++++++++++++++++++++++++++++++
\subsection{Robustness to variations of the rf field polarization}
\label{sec:rob2}
%--------------------------------------------------------------------------------------
\begin{figure}
\begin{center}
\resizebox{0.47\textwidth}{!}{\includegraphics{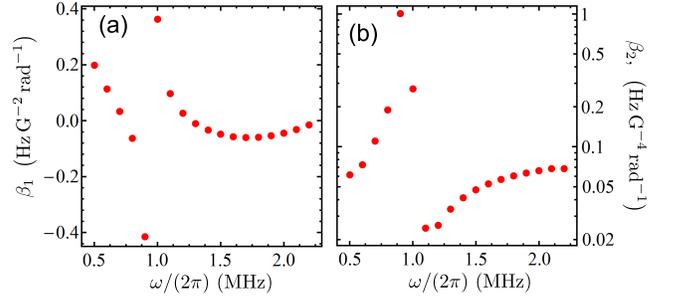}}
\end{center}
\caption{(color online) Coefficients $\beta_1$ (a) and $\beta_2$ (b) in the expansion (\ref{eq:19}), characterizing the robustness of the relative energy shift $\Delta E$  with respect to variations of the rf field polarization.}
\label{fig:7}
\end{figure}
%--------------------------------------------------------------------------------------- 

We parametrize the polarization of the rf field by the angle $\delta$, see expression~(\ref{eq:2}). $\delta = -\pi/4$ corresponds to perfect left-hand polarization, but in a physical implementation, $\delta$ may deviate from this value by an offset $\epsilon=\delta+\pi/4$.  
%--------------------------------------------------------------------------------------
\begin{figure*}[ht]
\begin{center}
\resizebox{0.75\textwidth}{!}{\includegraphics{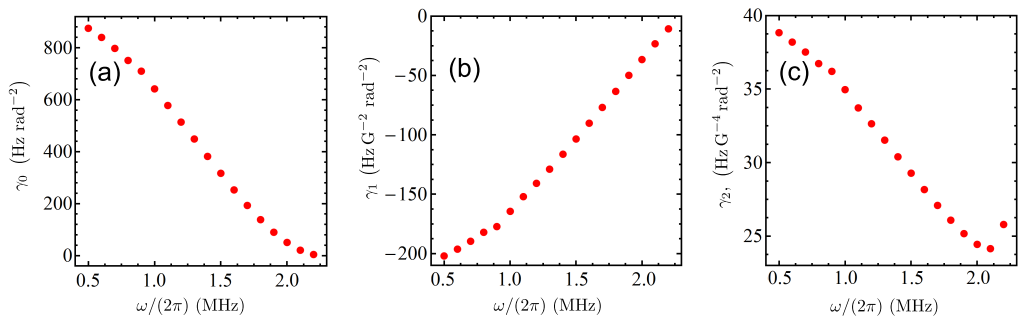}}
\end{center}
\caption{(color online) Parameters $\gamma_0$ (a), $\gamma_1$ (b), and $\gamma_2$ (c) in the expansion (\ref{eq:19}), characterizing the robustness of the relative energy shift $\Delta E$  with respect to variations of the rf field polarization.}
\label{fig:8}
\end{figure*}
%---------------------------------------------------------------------------------------
The local {dc} field can be characterized by a pair of angles $(\theta,\alpha)$, or equivalently by a pair $(\chi,\alpha)$, see Section~\ref{sec:mod1} for details. If the rf field polarization deviates from the perfectly circular one, energies of atomic states and hence $\Delta E$ experience an $\alpha$-dependent contribution. For reason of symmetry, $\Delta E(\alpha)=\Delta E(-\alpha)=\Delta E(\pi+\alpha)$, and for small $\epsilon$, the lowest-order harmonic, proportional to $\cos(2\alpha)$, gives the main contribution to the $\alpha$-dependent part. Also, an additional $\alpha$-independent contribution, quadratic in $\epsilon$ appears. As in the previous section, it is convenient to consider the expansion~(\ref{e:17}), and, in turn, expand coefficients $A_i$ as

%----------------------------------------------
\begin{align}
A_i(\epsilon,\alpha)&=A_i+\beta_i \cos(2\alpha) \epsilon +\gamma_{i} \epsilon^2+..., \label{eq:19}
\end{align}
%----------------------------------------------
where, again, $A_i=A_i(B_I^{m},B_{rf}^m,\delta=-\pi/4)$. It is easy to see that the coefficient $\beta_0=0$. The coefficients $\beta_1$ and $\beta_2$ are represented in Figure~\ref{fig:7}, coefficients $\gamma_0$, $\gamma_1$, and $\gamma_2$ are represented in Figure~\ref{fig:8}, 

%---------------------------------------------------------------------------------------
%+++++++++++++++++++++++++++++++++++++++++++++++++++++++++++++++++++++++++++++++++++++++
\subsection{Discussion}
\label{sec:rob3}
%--------------------------------------------------------------------------------------

We find that the behaviour of the coefficients $\alpha^{(I)}_{i}$, $\alpha^{(rf)}_{i}$, $\beta_i$, and $\gamma_i$ characterizing the response to fluctuations as well as the coefficient $A_3$ (see Figure~\ref{fig:4}(d)) show a qualitatively different dependence on the rf field frequency $\omega$. For example, the  values of $\alpha^{(I)}_0$ and $\alpha^{(rf)}_0$ go to zero when $\omega$ approaches the single-photon resonance, rendering the system more robust against fluctuations. However, at the same time,  $A_3$, describing the remaining energy inhomogeneity of the dressed trapping potential, grows. The optimal choice of the specific rf field frequency hence depends on the given instrumental stabilities of Ioffe and radio-frequency fields, on the deviation of the rf field polarization from the perfect left-hand circular one, and on the temperature of the atomic cloud.

\begin{figure}[b]
\begin{center}
\resizebox{0.46\textwidth}{!}{\includegraphics{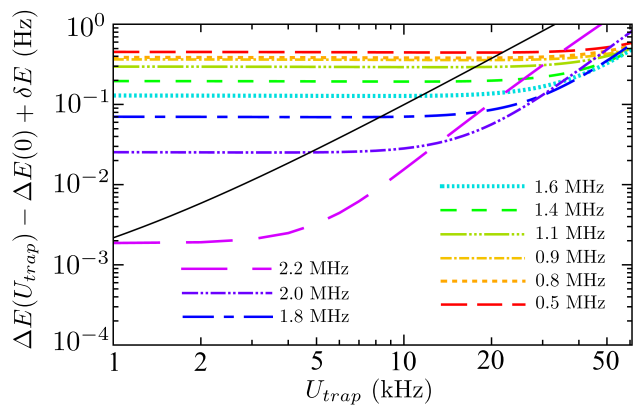}}
\end{center}
\caption{(color online) Relative energy shift including field and polarization deviations as described in the text for a {dc} first-order magic trap (black solid curve), and for rf dressed second-order magic traps (coloured non-continuous curves) as a function of $U_{trap}=V_{ad}(\rho)-V_{ad}(0)$. Frequencies of rf dressing fields $B_{rf}^m$ are given in the legend.}

\label{fig:9}
\end{figure}

As an example, we consider an atom chip setup with field deviations $\delta B_I/B_I=2.5\times 10^{-4}$ and $\delta B_{rf}/B_{rf}=5\times 10^{-4}$, the deviation of the rf field polarization from the perfect circular one can be estimated to $\epsilon=0.2^\circ$. Such parameters were recently realized in Ref~\cite{Berrada13}. Figure~\ref{fig:9} shows the value $\Delta E - \Delta E(0)+\delta E$ as a function of the trapping potential $U_{trap}(\chi)$ in the same manner as in Figure~\ref{fig:5}. Here we included a position-dependent mean square variation $\delta E$ of the relative energy shift:
%----------------------------------------------
\begin{equation*}
\delta E = \sqrt{\left(\frac{\partial \Delta E}{\partial B_I}\delta B_I\right)^2+\left(\frac{\partial \Delta E}{\partial B_{rf}}\delta B_I\right)^2+\left(\frac{\partial \Delta E}{\partial \epsilon}\epsilon\right)^2}.
\end{equation*}
%----------------------------------------------

%--------------------------------------------------------------------------------------
%---------------------------------------------------------------------------------------
Comparing Figure~\ref{fig:9} with Figure~\ref{fig:5}, one can see that the fluctuations only  weakly affect the undressed trap (black solid lines), but become more important for dressed ``second-order magic'' traps. For an atomic ensemble of a temperature of $1\,\mu$K (corresponding to $U_{trap}\simeq 20$\,kHz), the optimal rf frequency for ``second-order magic'' dressing lies between 1.8\,MHz and 2.2\,MHz. It still allows to increase the quality of the dressed potential by about one order of magnitude compared to the ``first-order magic'' condition. Further improvement may be possible when combining a static near-magic configuration similar to one presented in Figure~\ref{fig:f1}(c), blue dotted line, with rf dressing and taking into account the effect of atom interactions. A detailed optimization is beyond the scope of the present paper.
%%%%%%%%%%%%%%%%%%%%%%%%%%%%%%%%%%%%%%%%%%%%%%%%%%%%%%%%%%%%%%%%%%%%%%%%%%%%%%%%%%%%%%%%
%%%%%%%%%%%%%%%%%%%%%%%%%%%%%%%%%%%%%%%%%%%%%%%%%%%%%%%%%%%%%%%%%%%%%%%%%%%%%%%%%%%%%%%%
\section{Conclusion}
In conclusion, we propose rf dressing as a simple and flexible technique to suppress position-dependent dephasing of atomic ``clock'' superposition states in a magnetic Ioffe-Pritchard trap. For $^{87}$Rb, we have identified ``second-order magic'' conditions, where not only the first but also the second derivative of the relative energy shift with respect to the trapping potential vanishes. We have studied the robustness of these ``second-order magic'' conditions to deviations of the involved static and oscillating fields and find that for parameters realized in current atom chip experiments, the dressing can improve the quality of the trapping potential by about 1 order of magnitude compared to static ``first-order magic'' traps.

%%%%%%%%%%%%%%%%%%%%%%%%%%%%%%%%%%%%%%%%%%%%%%%%%%%%%%%%%%%%%%%%%%%%%%%%%%%%%%%%%%%%%%%%
%%%%%%%%%%%%%%%%%%%%%%%%%%%%%%%%%%%%%%%%%%%%%%%%%%%%%%%%%%%%%%%%%%%%%%%%%%%%%%%%%%%%%%%%
\section{Acknowledgements}
This work was supported by the Austrian Science Fund (FWF), project I 1602. 
%%%%%%%%%%%%%%%%%%%%%%%%%%%%%%%%%%%%%%%%%%%%%%%%%%%%%%%%%%%%%%%%%%%%%%%%%%%%%%%%%%%%%%%%
%%%%%%%%%%%%%%%%%%%%%%%%%%%%%%%%%%%%%%%%%%%%%%%%%%%%%%%%%%%%%%%%%%%%%%%%%%%%%%%%%%%%%%%%
\section*{APPENDIX}
Here we briefly review the Floquet theory following Ref.~\cite{Shirley65} and discuss the classification of the quasienergy spectrum within the weak-field Floquet approach. 

Firstly, Hamiltonians (\ref{eq:14}) are Hermitian matrices of periodic functions of $t$ with period $T=2\pi/\omega$. According the Floquet theorem, for a periodic Hamiltonian $\hat{H}(T)$, the Schr\"odinger equation
%----------------------------------------------------------
\begin{equation}
i\hbar \frac{\partial \Psi(t)}{\partial t}=\hat{H}(t) \Psi(t) \label{eq:20}
\end{equation}
%----------------------------------------------------------
has a fundamental matrix 
%----------------------------------------------------------
\begin{equation}
\FFF(t)=\left(\Psi_1(t),\Psi_2(t),...\right) \label{eq:21}
\end{equation}
%----------------------------------------------------------
which can be expressed in the form
%----------------------------------------------------------
\begin{equation}
\FFF(t)=\hat{\Phi}(t)\exp(-i\QQ t), \label{eq:22}
\end{equation}
%----------------------------------------------------------
where $\hat{\Phi}(t+T)=\hat{\Phi}(t)$ is a periodic matrix, and 
%----------------------------------------------------------
\begin{equation}
\QQ=\left(
\begin{array}{ccc}
q_1 & 0 & ... \\
0 & q_2 & ... \\
... & ...   & ...
\end{array}
\right) \label{eq:23}
\end{equation}
%----------------------------------------------------------
is a constant diagonal matrix. Values $\hbar q_i$ are called {\em quasienergies}. Note that these quasienergies are defined up to a shift by $n\hbar\omega$ corresponding to a change by $n$ in the number of photons describing the field responsible for the time-dependent terms in the Hamiltonian.

The matrix elements of $\FFF(t)$ can be written as
%----------------------------------------------------------
\begin{equation}
\FF_{\alpha \beta}(t)=\sum_n \FF_{\alpha \beta}^{(n)} \exp{[i(n\omega-q_\beta) t]}, \label{eq:24}
\end{equation}
%----------------------------------------------------------
and the Hamiltonian $\hat{H}$ can also be expanded into the Fourier series:
%----------------------------------------------------------
\begin{align}
\hat{H}(t)&=\sum_n \hat{H}^{(n)} \exp{[in\omega t]}, \quad \mathrm{or} \nonumber \\ 
H_{\alpha \beta}(t)&=\sum_n H_{\alpha \beta}^{(n)} \exp{[in\omega t]}. \label{eq:25}
\end{align}
%----------------------------------------------------------
The equation for the fundamental matrix
%----------------------------------------------------------
\begin{equation}
i\hbar \frac{\partial \FFF(t)}{\partial t}=\hat{H}(t) \FFF(t) \label{eq:26}
\end{equation}
%----------------------------------------------------------
can be rewritten using (\ref{eq:24}) and (\ref{eq:25}) as
%----------------------------------------------------------
\begin{equation}
\sum_{\gamma,m}\left(\frac{H_{\alpha \gamma}^{(k-m)}}{\hbar}
+k\omega \delta_{\alpha \gamma}\delta_{km} \right)\FF^{(m)}_{\gamma \beta}=
q_{\beta} \FF^{(k)}_{\alpha \beta}, \label{eq:27}
\end{equation}
%----------------------------------------------------------
where $\delta_{ij}$ is a Kronecker delta. Equation (\ref{eq:27}) can be written as infinite block matrix:
%----------------------------------------------------------
\begin{widetext}
\begin{equation}
\left(
\begin{array}{*7{c}}
... & ... & ... & ... & ... & ... & ...\vphantom{\dfrac{3}{3}} \\
%\\
... & \dfrac{\hat{H}^{(0)}}{\hbar}-2\omega\III &\dfrac{\hat{H}^{(-1)}}{\hbar}  & \dfrac{\hat{H}^{(-2)}}{\hbar} & ... & ... & ... \vphantom{\dfrac{\frac{3}{3}}{\frac{3}{3}}} \\
%\\
... & \dfrac{\hat{H}^{(1)}}{\hbar} & \dfrac{\hat{H}^{(0)}}{\hbar}-\omega\III & \dfrac{\hat{H}^{(-1)}}{\hbar} & \dfrac{\hat{H}^{(-2)}}{\hbar} & ... & ... \vphantom{\dfrac{\frac{3}{3}}{\frac{3}{3}}}\\
%\\
... & \dfrac{\hat{H}^{(2)}}{\hbar} & \dfrac{\hat{H}^{(1)}}{\hbar} & \hphantom{aa} \dfrac{\hat{H}^{(0)}}{\hbar}\hphantom{aa} & \dfrac{\hat{H}^{(-1)}}{\hbar} & \dfrac{\hat{H}^{(-2)}}{\hbar} & ... \vphantom{\dfrac{\frac{3}{3}}{\frac{3}{3}}}\\
%\\
... & ... & \dfrac{\hat{H}^{(2)}}{\hbar} & \dfrac{\hat{H}^{(1)}}{\hbar} & \dfrac{\hat{H}^{(0)}}{\hbar}+\omega\III & \dfrac{\hat{H}^{(-1)}}{\hbar} &  ... \vphantom{\dfrac{\frac{3}{3}}{\frac{3}{3}}}\\
%\\
... &... & ... & \dfrac{\hat{H}^{(2)}}{\hbar} & \dfrac{\hat{H}^{(1)}}{\hbar} & \dfrac{\hat{H}^{(0)}}{\hbar}+2 \omega\III & ...  \vphantom{\dfrac{\frac{3}{3}}{\frac{3}{3}}}\\
%\\
... & ... & ... & ... & ... & ... & ... \vphantom{\dfrac{3}{3}} \\
\end{array}
\right)
\left(
\begin{array}{{c}}
... \vphantom{\dfrac{3}{3}} \\
%\\
\FF_\beta^{(-2)} \vphantom{\dfrac{\frac{3}{3}}{\frac{3}{3}}} \\
%\\
\FF_\beta^{(-1)} \vphantom{\dfrac{\frac{3}{3}}{\frac{3}{3}}} \\
%\\
\FF_\beta^{(0)} \vphantom{\dfrac{\frac{3}{3}}{\frac{3}{3}}} \\
%\\
\FF_\beta^{(1)} \vphantom{\dfrac{\frac{3}{3}}{\frac{3}{3}}} \\
%\\
\FF_\beta^{(2)} \vphantom{\dfrac{\frac{3}{3}}{\frac{3}{3}}} \\
%\\
...  \vphantom{\dfrac{3}{3}} \\
\end{array}
\right)=q_\beta
\left(
\begin{array}{{c}}
... \vphantom{\dfrac{3}{3}} \\
%\\
\FF_\beta^{(-2)} \vphantom{\dfrac{\frac{3}{3}}{\frac{3}{3}}} \\
%\\
\FF_\beta^{(-1)} \vphantom{\dfrac{\frac{3}{3}}{\frac{3}{3}}} \\
%\\
\FF_\beta^{(0)} \vphantom{\dfrac{\frac{3}{3}}{\frac{3}{3}}} \\
%\\
\FF_\beta^{(1)} \vphantom{\dfrac{\frac{3}{3}}{\frac{3}{3}}} \\
%\\
\FF_\beta^{(2)} \vphantom{\dfrac{\frac{3}{3}}{\frac{3}{3}}} \\
%\\
...  \vphantom{\dfrac{3}{3}} \\
\end{array}
\right), \label{eq:28}
\end{equation}
\end{widetext}
%----------------------------------------------------------
where $\III$ is the identity matrix, and $\FF^{(n)}_{\beta}$ is $\beta$-th column of the matrix $\FFF^{(n)}$.

For practical calculations, ones truncates the equation~(\ref{eq:28}) to some finite number of Floquet blocks, in our simulation we used $21\times 21$ blocks. Note also, that within our weak-field Floquet approach, the main Fourier component $\hat{H}^{(0)}$ (\ref{eq:15}) of the Hamiltonian (\ref{eq:14}) is much larger than he non-zero frequency Fourier components $\hat{H}^{(-2)}$, $\hat{H}^{(-1)}$, $\hat{H}^{(1)}$ and $\hat{H}^{(2)}$, see (\ref{eq:16}). The rotating wave approximation consists in neglecting all of the non-zero frequency components, and equation (\ref{eq:28}) becomes a set of non-coupled matrix equations describing the atom-field system (up to a constant energy shift) in the semiclassical limit. In WFFA, these non-zero frequency terms are kept and responsible for couplings between different Floquet blocks, but they remain small. Therefore, for every $\omega$ in the range of interest (except the multiphoton resonances), eigenvectors of the Floquet ``Hamiltonian'' on the left side of the equation~(\ref{eq:28}) will have only small components everywhere except in some specific Floquet block. This allows to attribute the corresponding eigenvalue $q_{\beta}$ of the Floquet ``Hamiltonian'' to this Floquet block. 

If the set of equations~(\ref{eq:28}) is infinite, the quasienergy spectrum is periodic with period $\hbar \omega$ (which corresponds to different number of photons), but in a truncated set of equations used in practical calculations, this periodicity is not exact. Let us call ``true quasienergies'' the quasienergies $\hbar q_\beta$ which converges to the eigenvalues of the Hamiltonian $\hat{H}^{(0)}$ in the zero limit of all the $\hat{H}^{(n)}$ with $n\neq 0$. It is easy to see that these true quasienergies correspond to the central Floquet block. This classification method breaks down near the multiphoton resonances with level anticrossings, but everywhere else it can be applied and used for a numerical search for the second-order magic condition.

%%%%%%%%%%%%%%%%%%%%%%%%%%%%%%%%%%%%%%%%%%%%%%%%%%%%%%%%%%%%%%%%%%%%%%%%%%%%%%%%%%%%%%%%%
%%%%%%%%%%%%%%%%%%%%%%%%%%%%%%%%%%%%%%%%%%%%%%%%%%%%%%%%%%%%%%%%%%%%%%%%%%%%%%%%%%%%%%%%  


\begin{thebibliography}{99}

\bibitem{Bauch03} A.~Bauch, Meas. Sci. Technol. {\bf 14}, 1159–1173 (2003)

\bibitem{Santarelli99} G.~Santarelli, Ph.~Laurent, P.~Lemonde, A.~Clairon, A.~G.~Mann, S.~Chang, A.~N.~Luiten, C.~Salomon, Phys. Rev. Lett. {\bf 82}, 4619 (1999)

\bibitem{Rosenbusch09} P.~Rosenbusch,% {\em Magnetically trapped atoms for compact atomic clocks}. 
Appl. Phys. B {\bf 95}, 227–235 (2009)

\bibitem{Lacroute10} C.~Lacroute, F.~Ramirez-Martinez, P.~Rosenbusch, F.~Reinhard, C.~Deutsch, T.~Schneider, J.~Reichel, IEEE Trans. Ultrason. Ferroelectr. Freq. Control. {\bf 57}, 106-110 (2010)

\bibitem{Ramirez11} F.~Ram\'irez-Mart\'inez, C.~Lacro\^ute, P.~Rosenbusch, F.~Reinhard, C.~Deutsch, T.~Schneider, J.~Reichel, % {\em Compact frequency standard using atoms trapped on a chip}. 
Advances in space research {\bf 47}, 247-252 (2011)

\bibitem{Zobay01} O.~Zobay, B.~M.~Garraway, Phys. Rev. Lett. {\bf 86}, 1195 (2001)

\bibitem{Lesanovsky06} I.~Lesanovsky, T.~Schumm, S.~Hofferberth, L.~M.~Andersson, P.~Kr\"uger, and J.~Schmiedmayer, Phys. Rev. A {\bf 73}, 033619 (2006)

\bibitem{Schumm05} T.~Schumm {\em et al.,} Nature Physics {\bf 1}, 57-62 (2005)

\bibitem{Hofferberth07} S.~Hofferberth, B.~Fischer, T.~Schumm, J.~Schmiedmayer, I.~Lesanovsky, Phys. Rev. A {\bf 76}, 013401 (2007)

\bibitem{Merloti13} K.~Merloti, R.~Dubessy, L.~Longchambon, A.~Perrin, P.-E.~Pottie, V.~Lorent, and H.~Perrin, New J. Phys. {\bf 15}, 033007 (2013)

\bibitem{Fernholz07} T.~Fernholz, R.~Gerritsma, P.~Kr\"uger, and R.~J.~C.~Spreeuw, Phys. Rev. A {\bf 75}, 063406 (2007)

\bibitem{Sinuco13} G.~Sinuco-Le\'on, B.~M.~Garraway, New Journal of Physics {\bf 14}, 123008 (2012) 

\bibitem{Zanon12} T.~Zanon-Willette, E.~de Clercq, E.~Arimondo, Phys. Rev. Lett. {\bf 109}, 223003 (2012)

\bibitem{Jones13} L.~A.~Jones, J.~D.~Carter, J.~D.~D.~ Martin, Phys. Rev. A {\bf 87}, 023423 (2013)

\bibitem{Sarkany14} L.~S\'ark\'any, P.~Weiss, H.~Hattermann, J. Fort\'agh, Phys. Rev. A {\bf 90}, 053416 (2014)


\bibitem{Folman02} R.~Folman, P.~Kr\"uger, J.~Schmiedmayer, J.~Denschlang, C.~Henkel, Adv. At. Mol. Opt. Phys. {\bf 48}, 263-356 (2002) 

%\bibitem{Farkas10} D.~M.~Farkas, A.~Zozulya, D.~Z.~Anderson, {\em A compact microchip atomic clock based on all-optical interrogation of ultra-cold trapped Rb atoms}. Appl Phys B {\bf 101}, 705–721 (2010)

%\bibitem{Treutlein06} P.~Treutlein, T.~Steinmetz, Y.~Colombe, B.~Lev, P.~Hommelhoff,J.~Reichel, M.~Greiner, O.~Mandel, A.~Widera, T.~Rom, I.~Bloch, T.~W.~H\"ansch, {\em Quantum information processing in optical lattices and magnetic microtraps}. Fortschr. Phys. {\bf 54} 702–718 (2006)

\bibitem{Steck08} Daniel A. Steck, ``Rubidium 87 D Line Data,'' available online at http://steck.us/alkalidata (revision 2.1.4, 23 December 2010).

\bibitem{Shirley65} J.~H.~Shirley, Phys. Rev. {\bf 138}, B979-B987 (1965)


\bibitem{Chu04} S.-I. Chu, D.~A.~Telnov, Physics Reports {\bf 390} 1–131 (2004)

\bibitem{Berrada13} T.~Berrada, S.~Van~Frank, R.~B\"ucker, T.~Schumm, J.~F.~Schaff, J.~Schmiedmayer, Nature Communications {\bf 4}, 2077 (2013)











\end{thebibliography}
\end{document}